\documentclass[a4paper,aps,pre,floatfix,nofootinbib,twocolumn]{revtex4-1}

\usepackage{amsmath}
\usepackage{amssymb}
\usepackage[pdftex]{graphicx}
\usepackage{multirow}
\usepackage[english]{babel}

\begin{document}

\title{Who is the best player ever? A complex network analysis of the history of professional tennis}
\author{Filippo Radicchi}
\affiliation{Chemical and Biological Engineering, Northwestern University,
2145 Sheridan Road, Evanston, IL 60208, US}

\begin{abstract}
We consider all matches played by professional
tennis players between $1968$ and $2010$, and,
on the basis of this data set, 
construct a directed and weighted network of contacts. 
The resulting graph shows complex features, typical
of many real networked systems studied in literature. 
We develop a diffusion algorithm and apply it
to the tennis contact network in order
to rank professional players. {\it Jimmy Connors} is 
identified as the best player
of the history of tennis according to our
ranking procedure. 
We perform a complete analysis by determining 
the best players on specific playing surfaces
as well as the best ones in each of the years
covered by the data set.
The results of our technique are compared to those
of two other well established methods. 
In general, we observe that our ranking method
performs better: it has a higher predictive power
and does not require the arbitrary introduction
of external criteria for the correct assessment
of the quality of players. The present work
provides a novel evidence of the utility 
of tools and methods of network theory in real applications.
\end{abstract}

\maketitle

\section{Introduction}
Social systems generally display complex features~\cite{castellano09}.
Complexity is present at  the individual level: the behavior
of humans often obeys complex dynamical patterns
as for example demonstrated by the rules
governing electronic correspondence
~\cite{barabasi05,
malmgren08,radicchi09a,wu10}. At the same time,
complexity is present also at the global level.
This can be seen for example when social systems
are mathematically represented in terms
of graphs or networks, where vertices identify individuals
and edges stand for interactions between pairs of
social agents. Social networks are in most of the cases
{\it scale-free}~\cite{Barabasi1999}, indicating therefore a strong
degree of complexity from the topological and
global points of view. 
\\
During last years, the analysis of social systems has become
an important topic of interdisciplinary research and
as such has started to be not longer of interest to social scientists only.
The presence of a huge amount of digital data,
describing the activity of humans and the way in which
they interact, has made possible the analysis 
of large-scale systems. 
This new trend of research does not focus
on the behavior of single agents, but mainly
on the analysis of the macroscopic and statistical properties
of the whole population, with the aim to discover regularities and
universal rules. In this sense, professional
sports also represent optimal
sources of data. Soccer~\cite{onody04, duch10, heuer10},
football~\cite{girvan02, naim05}, baseball~\cite{petersen08,
sire09,saavedra09,Petersen2011} and basketball~\cite{naim07,skinner10}
are some remarkable cases in which network analysis
revealed features not visible with traditional approaches.
These are practical examples of the general outcome produced
by the intense research activity of last years: network
tools and theories do not serve only for descriptive
purposes, but have also wide practical applicability.
Representing a real system as a network
allows in fact to have a global view
of the system and simultaneously
use the entire information encoded by its complete list
of interactions. 
Particularly relevant results are those regarding:
the robustness of networks
under intentional attacks~\cite{albert00}; the spreading
of viruses in graphs~\cite{satorras01}; synchronization
processes~\cite{arenas08}, social models~\cite{castellano09},
and evolutionary and coevolutionary games~\cite{Szabo2007, perc10}
taking place on networks.
In this context fall also ranking techniques 
like the PageRank algorithm~\cite{Brin1},
where vertices are ranked on the basis of their ``centrality''
in a diffusion process occurring on the graph.
Diffusion algorithms, originally proposed for ranking
web pages, have been recently applied to citation networks~\cite{price65}.
The evaluation of the popularity of papers~\cite{chen07}, 
journals~\cite{west08,west10} and scientists~\cite{radicchi09}
is performed not by looking at local properties
of the network (i.e., number of citations)
but by measuring their degree of centrality in the flow
of information diffusing over the entire graph.
The use of the whole network leads to better evaluation
criteria without the addition of external
ingredients because the complexity of the citation process
is encoded by the topology of the graph.
\\
\begin{figure*}[!ht]
\begin{center}
\includegraphics[width=0.95\textwidth]{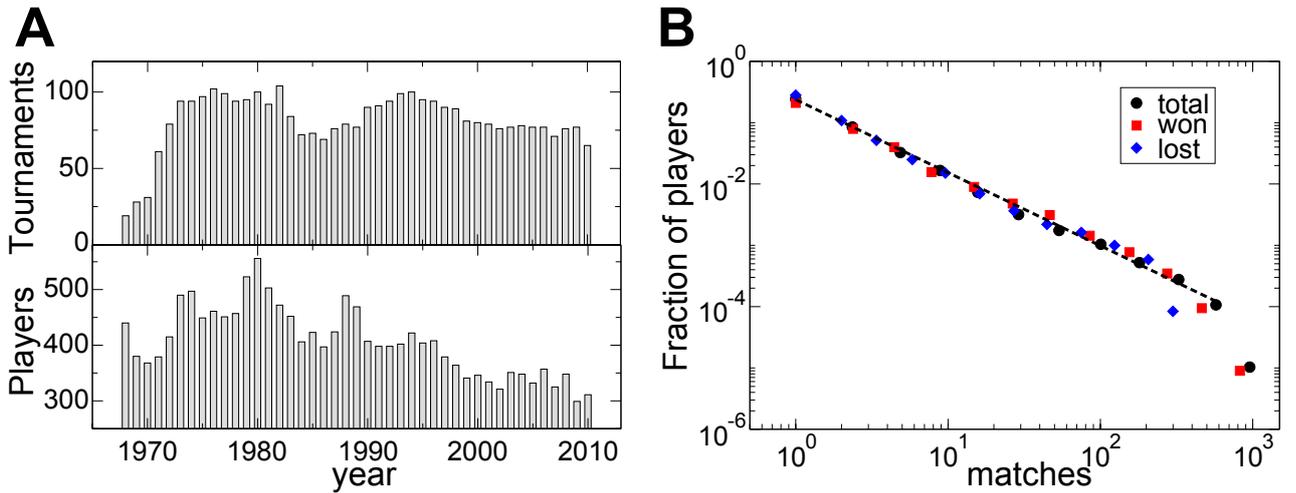}
\end{center}
\caption{Properties of the data set.
In panel a, 
we report the total number of tournaments (top panel) 
and players  (bottom panel) as a function of time. In panel b,
we plot the fraction of players having played (black circles), 
won (red squares) and lost (blue diamonds) a certain number of matches.
The black dashed line corresponds to the best power-law fit
with exponent consistent with the value $1.2(1)$.}
\label{fig1}
\end{figure*}
In this paper we continue in this direction of research
and present a novel example of a real system,
taken from the world of professional sports, suitable
for network representation. We consider the list of all 
tennis matches played by professional players during the last
$43$ years ($1968$-$2010$). Matches are considered
as basic contacts between the actors in the network
and weighted connections are drawn on the basis of the number
of matches between the same two opponents.
We first provide evidence of the complexity of the network
of contacts between tennis players. We then develop a ranking algorithm
similar to PageRank and quantify the importance
of tennis players with the so-called ``prestige score''.
The results presented here indicate once more that
ranking techniques based on networks outperform traditional methods.
The prestige score is in fact more accurate and has higher
predictive power than well established ranking schemes
adopted in professional tennis. More importantly,
our ranking method does not require the introduction
of external criteria for the assessment of the quality
of players and tournaments. Their importance is self-determined
by the various competitive processes 
described by the intricate network of contacts.
Our algorithm does nothing  more than taking into
account this information.

\section{Methods}

\subsection{Data set}

Data were collected from the
web site of the Association of Tennis
Professionals (ATP, {\tt www.atpworldtour.com}).
We automatically downloaded all matches played
by professional tennis players from January $1968$
to October $2010$. We restrict our analysis
only to matches played in Grand Slams and ATP World Tour tournaments
for a total of $3\,640$ tournaments and $133\,261$ matches.
For illustrative purposes, 
in the top plot of the panel a of Figure~\ref{fig1}, we report the
number of tournaments played in each of the years
covered by our data set. 
With the exception of the period between $1968$ and $1970$,
when ATP was still in its infancy, about $75$ tournaments were played each year.
Two periods of larger popularity  were registered 
around years $1980$ and $1992$ when
more than $90$ tournaments per year were played. 
The total number of different players present in our data set
is $3\,700$, and in the bottom plot of panel a of Figure~\ref{fig1}
we show how many players played at least one match in each 
of the years covered by our analysis. In this case,
the function is less regular. On average,
$400$ different players 
played in each of the years between $1968$ and $1996$.
Large fluctuations are anyway visible and 
a very high peak in $1980$,
when more than $500$ players participated in ATP tournaments,
is also present. Between $1996$ and $2000$, the number of players
decreased from $400$ to $300$ in an almost linear fashion. 
After that, the number of participants in ATP tournaments 
started to be more constant with small fluctuations 
around an average of about $300$ players.

\subsection{Network representation}

\begin{figure*}[!ht]
\begin{center}
\includegraphics[width=0.95\textwidth]{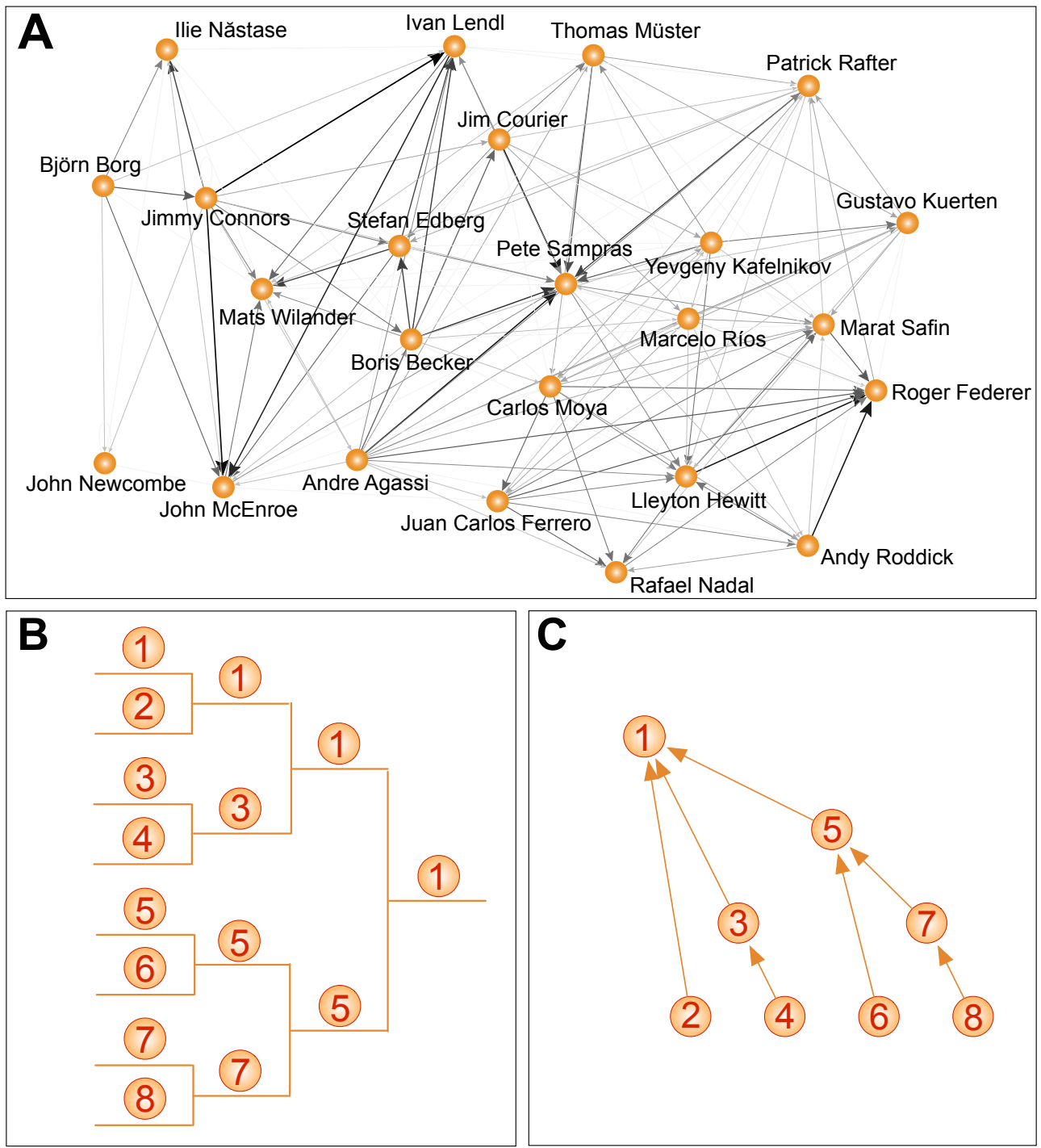}
\end{center}
\caption{Top player network and scheme for a single tournament. 
In panel a, we draw the subgraph of the contact network
restricted only to those players who have been number
one in the ATP ranking. Intensities and widths
are proportional to the logarithm of the weight
carried by each directed edge.
In panel b, we report a schematic view of the matches
played during a single tournament, while in panel c 
we draw the network derived from it.}
\label{fig2}
\end{figure*}

We represent the data set as a network of
contacts between tennis players. This is a very natural
representation of the system since a single match
can be viewed as an elementary contact between two opponents.
Each time the player $i$ plays
and wins against player $j$, we draw a directed
connection from $j$ to $i$ [$j \to i$, see Figure~\ref{fig2}]. 
We adopt a weighted representation of the contacts~\cite{Barrat2004}, 
by assigning to the generic directed edge $j \to i$ 
a weight $w_{ji}$ equal to the number of times that player $j$ looses
against player $i$. Our data are flexible and
allow various levels of representation
by including for example only matches
played in a certain period of time, on a certain type of surface, etc.
An example is reported in panel a of Figure~\ref{fig2} where
the network of contacts is restricted only to the
$24$ players having been number one in the
official ATP ranking. 
In general, networks obtained from the aggregation
of a sufficiently high number of matches have topological complex 
features consistent with the majority
of networked social systems so far studied
in literature~\cite{Reka:RevModPhys2002, Newman2003}. 
Typical measures revealing complex
structure are represented by the probability density
functions of the in- and out-strengths of vertices~\cite{Barrat2004},
both following a clear power-law behavior [see Figure~\ref{fig1}, panel b].
In our social system, this means
that most of the players perform a small number of 
matches (won or lost) and then quit playing in major tournaments.
On the other hand, a small set of top players performs many matches
against worse opponents (generally beating them) and also
many matches (won or lost) against other top players.
This picture is consistent with the so-called 
``Matthew effect'' in career longevity recently observed also
in other professional sports~\cite{petersen08,Petersen2011}.

\subsection{Prestige score}
The network representation can be used for ranking players. 
In our interpretation, each player in the network carries a unit of
``tennis prestige'' and we imagine that prestige flows in the graph 
along its weighted connections. The process can be mathematically
solved by determining the solution of the system of equations
\begin{equation}
P_i \; = \; \left(1-q\right) \, \sum_j \, P_j \, \frac{w_{ji}}{s_j^{out}} \; + \;
\frac{q}{N} \; + \;
\frac{1-q}{N} \, \sum_j \, P_j \, \delta\left(s_j^{out}\right)  \;\; , 
\label{eq:pg}
\end{equation}
valid for all nodes $i=1,\ldots,N$, with
the additional constraint that
$\sum_i P_i=1$. $N$ indicates
the total number of players (vertices) in the network, while
$s_j^{out} = \sum_{i} \, w_{ji}$ is the out-strength of the node
$j$ (i.e., the sum of the weight
of all edges departing from vertex $j$). $P_i$ is the ``prestige score''
assigned to player $i$ and represents
the fraction of the overall tennis prestige sitting,
in the steady state of the diffusion process, on vertex $i$. 
In Eqs.~(\ref{eq:pg}), $q \in \left[0,1\right]$ 
is a control parameter which accounts for the importance of the various 
terms contributing to the  score of the nodes.
The term $ \left(1-q\right) \, \sum_j \, P_j \, \frac{w_{ji}}{s_j^{out}}$ 
represents the portion of score received by node $i$ in the
diffusion process: vertices redistribute their entire credit 
to neighboring nodes proportionally
to the weight of the connections linking to them. $\frac{q}{N}$
stands for a uniform redistribution of tennis prestige
among all nodes according to which each player in the graph
receives a constant and equal amount of credit. Finally the term
$\frac{1-q}{N} \, \sum_j \, P_j \, \delta\left(s_j^{out}\right)$
[with $\delta\left(\cdot\right)$ equal
to one only if its argument is equal to zero, and
zero otherwise]  serves
as a correction in the case of existence of dandling nodes
(i.e., nodes with null out-strength), 
which otherwise would behave as sinks in the diffusion process.
Our prestige score is analogous to the PageRank score~\cite{Brin1},
originally formulated for ranking web pages and more recently
applied in different contexts.
\\
In general topologies, analytical solutions of Eqs.~(\ref{eq:pg}) 
are hard to find. The stationary values of the scores
$P_i$s can be anyway computed recursively, 
by setting at the beginning $P_i=1/N$ (but
the results do not depend on the choice of the initial value)
and iterating Eqs.~(\ref{eq:pg}) until they converge to values stable
within {\it a priori} fixed precision.

\subsubsection{Single tournament}

\begin{figure}[!ht]
\begin{center}
\includegraphics[width=0.45\textwidth]{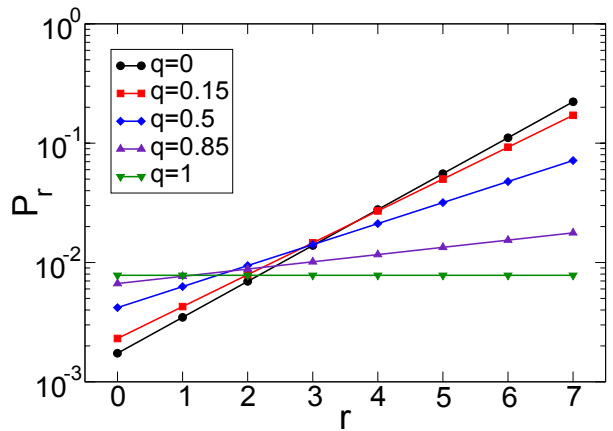}
\end{center}
\caption{Prestige score in a single tournament.
Prestige score $P_r$ as a function
of the number of victories $r$ in a tournament
with $\ell =7$ rounds (Grand Slam).
Black circles are obtained from Eqs.~(\ref{eq:s6}) and
valid for $q=0$. All other values of $q>0$ have
been calculated from Eqs.~(\ref{eq:s5}): red squares
stand for $q=0.15$, blue diamonds for $q=0.5$, violet
up-triangles for $q=0.85$ and green down-triangles for $q=1$.}
\label{fig3}
\end{figure}

In the simplest case in which the graph is obtained by aggregating 
matches of a single tournament only, we can analytically determine 
the solutions of Eqs.~(\ref{eq:pg}).
In a single tournament, matches are hierarchically
organized in a binary rooted tree and the topology of the resulting
contact network is very simple [see Figure~\ref{fig2}, panels b and c]. 
Indicate with $\ell$ the number of matches that the winner of 
the tournament should play (and win). The total number of players present
at the beginning of the tournament is $N=2^\ell$. The prestige score
is simply a function of $r$, the number of matches 
won by a player, and can be denoted by $P_r$.
We can rewrite Eqs.~(\ref{eq:pg}) as
\begin{equation}
P_r = P_0 \; + \; \left(1-q\right) \, \sum_{v=1}^{r} P_{v-1} \;\; ,  
\label{eq:s1}
\end{equation}
where $P_0=\frac{1-q}{N} \, P_{\ell} \; + \; \frac{q}{N}$
and  $0\leq r \leq \ell$.
The score $P_r$ is given by the sum
of two terms: $P_0$ stands  for the equal contribution
shared by all players independently of the number of victories;
$\left(1-q\right) \, \sum_{v=1}^{r} P_{v-1}$ represents
the score accrued for the number of matches won.
The former system of equations has a recursive solution
given by
\begin{equation}
P_r \; = \; \left(2-q\right) P_{r-1} \; = \ldots \; = \; \left(2-q\right)^r \, P_0 \;\; ,
\label{eq:s2}
\end{equation}
which is still dependent on a constant that can be determined by
implementing the normalization condition
\begin{equation}
\sum_{r=0}^\ell \, n_r \, P_r = 1 \;\;. 
\label{eq:s3}
\end{equation}
In Eq.~(\ref{eq:s3}), $n_r$ indicates the number of players 
who have won $r$ matches.
We have $n_r = 2^{\ell-r-1}$ for $0 \leq r < \ell$ and 
$n_\ell=1$ and Eqs.~(\ref{eq:s2}) and~(\ref{eq:s3}) allow to compute
\[
\begin{array}{ccc}
\left(P_0\right)^{-1}  = & \sum_{r=0}^{\ell-1} \, \left(2-q\right)^r \, 2^{\ell-1-r}  & + \; \left(2-q\right)^\ell
\\
& 2^{\ell-1} \, \sum_{r=0}^{\ell-1} \, \left( \frac{2-q}{2} \right)^r & + \; \left(2-q\right)^\ell \\
& 2^{\ell-1} \, \frac{1- \left[\left(2-q\right)/2\right]^\ell}{\left[\left(2-q\right)/2\right]} & + \; \left(2-q\right)^\ell\\ 
& \frac{2^\ell-\left(2-q\right)^\ell}{q} & + \; \left(2-q\right)^\ell
\end{array} \;\; .
\]
In the former calculations, we have used the 
well known identity 
$\sum_{r=0}^{v} \, x^r \, = \, \frac{1-x^{v+1}}{1-x}$, 
valid for any $\left|x\right|<1$ and $v\geq0$, which respectively
means $0 < q \leq 1$  and  $\ell>0$ in our case.  Finally, we obtain
\begin{equation}
P_0 \; = \; \frac{q}{2^\ell \, + \, \left(2-q\right)^\ell \, \left(q-1\right)} \;\; ,
\label{eq:s4}
\end{equation}
which together with Eqs.~(\ref{eq:s2}) provides the solution
\begin{equation}
P_r \; = \; \frac{q \; \left(2-q\right)^{r}}{2^\ell \, + \, \left(2-q\right)^\ell \, \left(q-1\right)} \;\; .
\label{eq:s5}
\end{equation}
It is worth to notice that for $q=1$, Eqs.~(\ref{eq:s5}) 
correctly give $P_r=2^{-\ell}$ for any $r$,
meaning that, in absence of
diffusion, prestige is homogeneously distributed among all nodes. 
Conversely, for $q=0$ the solution  is
 \begin{equation}
P_r \; = \; \frac{2^{r}}{2^{\ell-1} \, \left(\ell+2\right)} \;\; .
\label{eq:s6}
\end{equation}
In Figure~\ref{fig3}, we plot Eqs.~(\ref{eq:s5}) and~(\ref{eq:s6})
for various values of $q$. In general, sufficiently
low values of $q$ allow to assign to the winner
of the tournament a score which is about two order of magnitude 
larger than the one given to players loosing at the first round.
The score of the winner is an exponential
function of $\ell$, the length of the tournament. Grand
Slams have for instance length $\ell=7$ and their
relative importance is therefore two or four times
larger than the one of other ATP tournaments, typically
having lengths $\ell=6$ or $\ell=5$.

\section{Results} 

\begingroup
\squeezetable
\begin{table}[!ht]
\begin{center}
\begin{tabular}{l l l l l l}
Rank & Player & Country & Hand & Start & End\\
\hline
\hline
$ 1$ & {\bf Jimmy Connors} & United States & L &  $ 1970$ &   $1996$ \\ 
 \hline 
$ 2$ & {\bf Ivan Lendl} &  United States &  R & $ 1978$ & $ 1994$ \\ 
 \hline 
$ 3$ & {\bf John McEnroe} &  United States &  L & $ 1976$ & $ 1994$ \\ 
 \hline 
$ 4$ & Guillermo Vilas & Argentina & L & $ 1969$ & $ 1992$ \\ 
 \hline 
$ 5$ & {\bf Andre Agassi} &  United States &  R & $ 1986$ & $ 2006$ \\ 
 \hline 
$ 6$ & {\bf Stefan Edberg} &  Sweden &  R & $ 1982$ & $ 1996$ \\ 
 \hline 
$ 7$ & {\bf Roger Federer} &  Switzerland &  R & $ 1998$ & $ 2010$ \\ 
 \hline 
$ 8$ & {\bf Pete Sampras} &  United States &  R & $ 1988$ & $ 2002$ \\ 
 \hline 
$ 9$ & {\bf Ilie N\u{a}stase} &  Romania &  R & $ 1968$ & $ 1985$ \\ 
 \hline 
$ 10$ & {\bf Bj\"orn Borg} &  Sweden &  R & $ 1971$ & $ 1993$ \\ 
 \hline 
$ 11$ & {\bf Boris Becker} &  Germany &  R & $ 1983$ & $ 1999$ \\ 
 \hline 
$ 12$ & Arthur Ashe & United States & R & $ 1968$ & $ 1979$ \\ 
 \hline 
$ 13$ & Brian Gottfried & United States & R & $ 1970$ & $ 1984$ \\ 
 \hline 
$ 14$ & Stan Smith & United States & R & $ 1968$ & $ 1985$ \\ 
 \hline 
$ 15$ & Manuel Orantes & Spain & L & $ 1968$ & $ 1984$ \\ 
 \hline 
$ 16$ & Michael Chang & United States & R & $ 1987$ & $ 2003$ \\ 
 \hline 
$ 17$ & Roscoe Tanner & United States & L & $ 1969$ & $ 1985$ \\ 
 \hline 
$ 18$ & Eddie Dibbs & United States & R & $ 1971$ & $ 1984$ \\ 
 \hline 
$ 19$ & Harold Solomon & United States & R & $ 1971$ & $ 1991$ \\ 
 \hline 
$ 20$ & Tom Okker & Netherlands & R & $ 1968$ & $ 1981$ \\ 
 \hline 
$ 21$ & {\bf Mats Wilander} &  Sweden &  R & $ 1980$ & $ 1996$ \\ 
 \hline 
$ 22$ & Goran Ivani\v{s}evi\'c & Croatia & L & $ 1988$ & $ 2004$ \\ 
 \hline 
$ 23$ & Vitas Gerulaitis & United States & R & $ 1971$ & $ 1986$ \\ 
 \hline 
$ 24$ & {\bf Rafael Nadal} &  Spain &  L & $ 2002$ & $ 2010$ \\ 
 \hline 
$ 25$ & Ra\'ul Ramirez & Mexico & R & $ 1970$ & $ 1983$ \\ 
 \hline 
$ 26$ & {\bf John Newcombe} &  Australia &  R & $ 1968$ & $ 1981$ \\ 
 \hline 
$ 27$ & Ken Rosewall & Australia & R & $ 1968$ & $ 1980$ \\ 
 \hline 
$ 28$ & {\bf Yevgeny Kafelnikov} &  Russian Federation &  R & $ 1992$ & $ 2003$ \\ 
 \hline 
$ 29$ & {\bf Andy Roddick} &  United States &  R & $ 2000$ & $ 2010$ \\ 
 \hline 
$ 30$ & {\bf Thomas Muster} &  Austria &  L & $ 1984$ & $ 1999$ \\ 
\hline
\hline
\end{tabular}
\end{center}
\caption{Top $30$ players in the history of tennis.  
From left
to right we indicate for each player: rank position according
to prestige score, full name, country
of origin, the hand used to play, and the years of the first
and last ATP tournament played. Players having been
at the top of ATP ranking are reported in boldface.}
\label{table1}
\end{table}
\endgroup

\begin{figure*}[!ht]
\begin{center}
\includegraphics[width=0.95\textwidth]{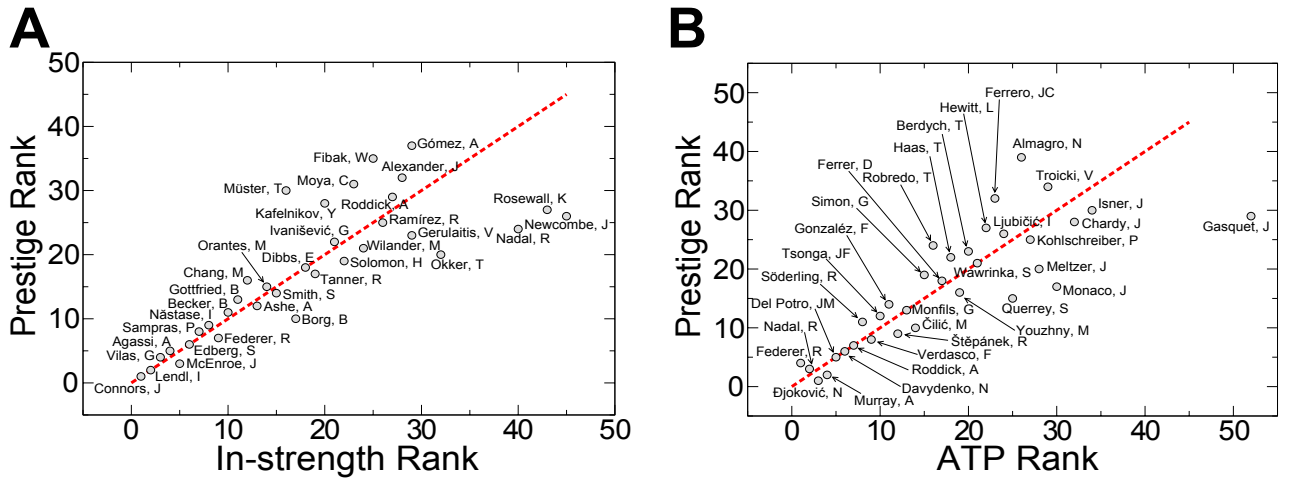}
\end{center}
\caption{Relation between prestige rank and other
ranking techniques.  In panel a, we present a
scatter plot of the prestige rank {\it versus} the rank based
on the number of victories (i.e., in-strength). Only players ranked
in the top $30$ positions in one of the two lists are reported.
Rank positions are calculated on the network corresponding
to all matches played between $1968$ and $2010$.
In panel b, a similar scatter plot is presented, but now
only matches of year $2009$ are considered for the construction
of the network. Prestige rank positions are compared with those
assigned by ATP.}
\label{fig4}
\end{figure*}

\begin{table}
\begin{center}
\begin{tabular}{l l l l}
Year & Prestige & ATP year-end & ITF\\
\hline \hline
$ 1968$ & Rod Laver &  - & -\\ 
 \hline 
$ 1969$ & Rod Laver & - & - \\ 
 \hline 
$ 1970$ & Rod Laver &  - & - \\ 
 \hline 
$ 1971$ & Ken Rosewall &  - & - \\ 
 \hline 
$ 1972$ & Ilie N\u{a}stase & - & - \\ 
 \hline 
$ 1973$ & Tom Okker & Ilie N\u{a}stase & - \\ 
 \hline 
$ 1974$ & Bj\"orn Borg & Jimmy Connors & -\\ 
 \hline 
$ 1975$ & Arthur Ashe & Jimmy Connors & - \\ 
 \hline 
$ 1976$ & Jimmy Connors &  Jimmy Connors & - \\ 
 \hline 
$ 1977$ & Guillermo Vilas & Jimmy Connors & - \\ 
 \hline 
$ 1978$ & Bj\"orn Borg & Jimmy Connors & Bj\"orn Borg\\ 
 \hline 
$ 1979$ & Bj\"orn Borg & Bj\"orn Borg &  Bj\"orn Borg\\ 
 \hline 
$ 1980$ & John McEnroe &  Bj\"orn Borg & Bj\"orn Borg\\ 
 \hline 
$ 1981$ & Ivan Lendl &  John McEnroe & John McEnroe\\ 
 \hline 
$ 1982$ & Ivan Lendl & John McEnroe & Jimmy Connors\\ 
 \hline 
$ 1983$ & Ivan Lendl &  John McEnroe & John McEnroe\\ 
 \hline 
$ 1984$ & Ivan Lendl &  John McEnroe & John McEnroe\\ 
 \hline 
$ 1985$ & Ivan Lendl &  Ivan Lendl & Ivan Lendl\\ 
 \hline 
$ 1986$ & Ivan Lendl &  Ivan Lendl & Ivan Lendl\\ 
 \hline 
$ 1987$ & Stefan Edberg &  Ivan Lendl & Ivan Lendl\\ 
 \hline 
$ 1988$ & Mats Wilander &  Mats Wilander & Mats Wilander\\ 
 \hline 
$ 1989$ & Ivan Lendl &  Ivan Lendl & Boris Becker\\ 
 \hline 
$ 1990$ & Stefan Edberg &  Stefan Edberg & Ivan Lendl\\ 
 \hline 
$ 1991$ & Stefan Edberg &  Stefan Edberg & Stefan Edberg\\ 
 \hline 
$ 1992$ & Pete Sampras &  Jim Courier & Jim Courier\\ 
 \hline 
$ 1993$ & Pete Sampras &  Pete Sampras & Pete Sampras\\ 
 \hline 
$ 1994$ & Pete Sampras &  Pete Sampras & Pete Sampras\\ 
 \hline 
$ 1995$ & Pete Sampras &  Pete Sampras & Pete Sampras\\ 
 \hline 
$ 1996$ & Goran Ivani\v{s}evi\'c &  Pete Sampras & Pete Sampras\\ 
 \hline 
$ 1997$ & Patrick Rafter &  Pete Sampras & Pete Sampras\\ 
 \hline 
$ 1998$ & Marcelo R\'ios &  Pete Sampras & Pete Sampras\\ 
 \hline 
$ 1999$ & Andre Agassi &  Andre Agassi & Andre Agassi\\ 
 \hline 
$ 2000$ & Marat Safin &  Gustavo Kuerten & Gustavo Kuerten\\ 
 \hline 
$ 2001$ & Lleyton Hewitt &   Lleyton Hewitt & Lleyton Hewitt\\ 
 \hline 
$ 2002$ & Lleyton Hewitt &   Lleyton Hewitt & Lleyton Hewitt\\ 
 \hline 
$ 2003$ & Roger Federer &  Andy Roddick & Andy Roddick\\ 
 \hline 
$ 2004$ & Roger Federer &  Roger Federer & Roger Federer\\ 
 \hline 
$ 2005$ & Roger Federer &  Roger Federer & Roger Federer\\ 
 \hline 
$ 2006$ & Roger Federer &  Roger Federer & Roger Federer\\ 
 \hline 
$ 2007$ & Rafael Nadal &  Roger Federer & Roger Federer\\ 
 \hline 
$ 2008$ & Rafael Nadal &  Rafael Nadal & Rafael Nadal\\ 
 \hline 
$ 2009$ & Novak Djokovi\'c &  Roger Federer & Roger Federer\\ 
 \hline 
$ 2010$ & Rafael Nadal &   Rafael Nadal & Rafael Nadal\\ 
 \hline 
\hline
\end{tabular}
\end{center}
\caption{Best players of the year.
For each year we report the best
player according to our ranking scheme and those of ATP and
ITF. Best year-end ATP players are listed for all  years from $1973$ on. ITF
world champions have started to be nominated
since $1978$ only.}
\label{table2}
\end{table}

We set $q=0.15$ and run the ranking procedure
on several networks derived from our data set.
The choice $q=0.15$ is mainly due to
tradition. This is the value originally used
in the PageRank algorithm~\cite{Brin1} and then
adopted in the majority of papers about this type of
ranking procedures~\cite{chen07,radicchi09,west08,west10}.
It should be stressed that $q=0.15$ is also a reasonable
value because it ensures a high relative
score for the winner of the tournament as stated
in Eqs.~(\ref{eq:s5}).
\\
In Table~\ref{table1}, we report the results 
obtained from the analysis of the contact network
constructed over the whole data set. The method
is very effective in finding the best players
of the history of tennis. In our top $10$ list,
there are $9$ players having been 
number one in the ATP ranking.
Our ranking technique identifies {\it Jimmy Connors}
as the best player of the history of tennis. 
This could be {\it a posteriori} justified by 
the extremely long and successful 
career of this player. Among all top players
in the history of tennis, {\it Jimmy Connors}
has been undoubtedly the one with the longest and most
regular trend, being in the top $10$ of 
the ATP year-end ranking for $16$ consecutive
years ($1973$-$1988$). Prestige score is strongly correlated
with the number of victories, but important
differences are evident when the two techniques are compared.
Panel a of Figure~\ref{fig4} shows a scatter plot,
where the rank calculated according to our score is
compared to the one based on the number of victories. An important
outlier is this plot is represented by the {\it Rafael Nadal},
the actual number one of the ATP ranking. {\it Rafael Nadal}
occupies the rank position number $40$ according to the
number of victories obtained in his still young career, but he is
placed at position number $24$ according to prestige score,
consistently with his high relevance in the recent
history of tennis.
A similar effect is also visible
for {\it Bj\"orn Borg}, whose career length
was shorter than average. He is ranked at position $17$ 
according to the number of victories. Prestige score
differently is able to determine the undoubted importance of 
this player and, in
our ranking, he is placed among the
best $10$ players of the whole history of 
professional tennis.
\\
In general, players still in activity are penalized with
respect to those who have ended their careers.
Prestige score is in fact strongly
correlated with the number of victories [see 
panel a of Figure~\ref{fig4}]
and still active players
did not yet played all matches of their career. 
This bias, introduced by the incompleteness of
the data set, can be suppressed by considering, for example,
only matches played in the same year.
Table~\ref{table2} shows the list
of the best players of the year according
to prestige score. It is interesting
to see how our score is effective
also here. We identify {\it Rod Laver} as the best tennis player
between $1968$ and $1971$, period
in which no ATP ranking was still established. 
Similar long periods of dominance are also those of 
{\it Ivan Lendl} ($1981-1986$),
{\it Pete Sampras} ($1992-1995$) and {\it Roger Federer}
 ($2003-2006$). For comparison, 
we report the best players of the 
year according to ATP (year-end rank)
and ITF (International Tennis Federation, {\tt www.itftennis.com})
rankings. In many cases, the best players of the year
are the same in all lists. 
Prestige rank seems however to have
a higher predictive power by anticipating
the best player of the subsequent year according
to the two other rankings.
{\it John McEnroe}
is the top player in our ranking in $1980$ and occupies the same
position in the ATP and ITF lists one year later.
The same happens also for {\it Ivan Lendl}, {\it Pete Sampras},
{\it Roger Federer} and {\it Rafael Nadal}, respectively best
players of the years $1984$, $1992$, $2003$ and $2007$ 
according to prestige score, but only
one year later placed at the top position of ATP and ITF rankings.
The official ATP rank and the one determined on the basis
of the prestige score are strongly correlated, but
small differences between them are very interesting.
An example is reported in  panel b of
Figure~\ref{fig4}, where the prestige rank 
calculated over the contact network of $2009$ is compared
with the ATP rank of the end of the same year 
(official ATP year-end rank as of December $28$, $2009$). The top $4$ positions
according to prestige score do not corresponds to those of the
ATP ranking. The best player of the year, for example, is
 {\it Novak Djokovi\'c} instead of {\it Roger Federer}.
\\
We perform also a different kind of analysis by constructing
networks of contacts for decades
and for specific types of playing surfaces.
According to our score, the best players per decade
are [Table~\ref{table3} lists the top $30$ players in each decade]
: {\it Jimmy Connors} ($1971-1980$), 
{\it Ivan Lendl} ($1981-1990$), 
{\it Pete Sampras} ($1991-2000$) and {\it Roger
Federer} ($2001-2010$). Prestige score identifies
 {\it Guillermo Vilas} as the best player ever in
clay tournaments, while
on grass and hard surfaces the best players
ever are {\it Jimmy Connors}
and {\it Andre Agassi}, respectively [see Table~\ref{table4} for
the list of the top $30$ players of a particular playing surface].

\section{Discussion}
Tools and techniques of complex networks
have wide applicability since many real
systems can be naturally described as graphs. 
For instance, rankings  based on diffusion
are  very effective since the whole information encoded by the network
topology can be used in place of simple local properties or
pre-determined and arbitrary criteria.
Diffusion algorithms, like the one for calculating 
the PageRank score~\cite{Brin1}, 
were first developed for ranking web pages and more recently have been
applied to citation networks~\cite{chen07,
radicchi09,west08,west10}. In citation networks, diffusion
algorithms generally outperform simple ranking techniques
based on local network properties (i.e., number of
citations). When the popularity
of papers is in fact measured in terms of mere citation
counts, there is no distinction between the
quality of the citations received. In contrast, when
a diffusion algorithm is used
for the assessment of the quality
of scientific publications, then it is not only important 
that popular papers receive many citations, but also that they are 
cited by other popular articles.
In the case of citation networks however,
possible biases are introduced in the absence of a proper
classification of papers in
scientific disciplines\cite{radicchi08}.
The average number of publications and citations strongly depend
on the popularity of a particular topic of research and
this fact influences the outcome of
a diffusion ranking algorithm. Another important issue
in paper citation networks is related to their intrinsic temporal nature:
connections go only backward in time,
because papers can cite only older articles and not {\it vice versa}.
The anisotropy of the underlying network  automatically
biases any method based on diffusion.
Possible corrections can be implemented:
for example, the weight of citations may be
represented by an exponential
decaying function of the age difference between citing
and cited papers~\cite{chen07}. 
Though these corrections can be reasonable, they
are {\it ad hoc} recipes and as such may be considered arbitrary.
\\
Here we have reported another emblematic 
example of a real social system
suitable for network representation:
the graph of contacts (i.e., matches) between professional
tennis players.  This network shows complex topological features
and as such the understanding of the whole system cannot be
achieved by decomposing the graph and studying each component in
isolation. In particular, the correct assessment of players' performances 
needs the simultaneously consideration of
the whole network of interactions. We have therefore introduced a 
new score, called ``prestige score'', based on a diffusion
process occurring on the entire network of contacts between tennis players.
According to our ranking technique, the relevance of players is not 
related to the number of victories only but mostly to the 
quality of these victories. In this sense, it could be more important
to beat a great player than to win many matches against less 
relevant opponents. The results of the analysis have revealed that 
our technique is effective in finding the best players of 
the history of tennis. The biases mentioned in the case of citation
networks are not present in the tennis contact
graph. Players do not need to be classified
since everybody has the opportunity
to participate to every tournament. Additionally, 
there is not temporal dependence
because matches are played between opponents still in activity
and the flow does not necessarily go from young players towards
older ones. In general, players still in activity
are penalized with respect to those who already
ended their career only for incompleteness of information
(i.e., they did not play all matches of their career) and
not because of an intrinsic bias of the system.  
Our ranking technique is furthermore effective
because it does not require any external criteria 
of judgment. As term of comparison,
the actual ATP ranking is based on the amount of
points collected by players during the season.
Each tournament has an {\it a priori} fixed value and points
are distributed accordingly to the round reached
in the tournament. In our approach differently, 
the importance of a tournament is self-determined: its 
quality  is established by  the level of the players who are taking part of it.
\\
In conclusion, we would like to stress that the aim of our method 
is not to replace other ranking techniques, optimized and 
almost perfected in the course of many years.
Prestige rank represents only a novel method with a 
different spirit and may be used  to corroborate the accuracy of other 
well established ranking techniques.

\begin{acknowledgments}
We thank the Association of Tennis Professionals for
making publicly available the data set of all tennis 
matches played during last $43$ years.  
Helpful discussions with Patrick McMullen are gratefully 
acknowledged as well.
\end{acknowledgments}

\begin{turnpage}
\begin{table*}
\begin{center}
\begin{tabular}{l |l l|l l|l l|l l|}
\cline{2-9}
& \multicolumn{2}{|c|}{{\normalsize $1971-1980$}} & \multicolumn{2}{|c|}{{\normalsize $1981-1990$}} & \multicolumn{2}{|c|}{{\normalsize $1991-2000$}} & \multicolumn{2}{|c|}{{\normalsize $2001-2010$}}\\
Rank & Player & Country & Player & Country & Player & Country & Player & Country\\
\hline\hline
$ 1$ & Jimmy Connors & United States & Ivan Lendl & United States & Pete Sampras & United States & Roger Federer & Switzerland\\ 
 \hline 
$ 2$ & Bj\"orn Borg & Sweden &  John McEnroe & United States & Andre Agassi & United States & Rafael Nadal & Spain\\ 
 \hline 
$ 3$ & Ilie N\u{a}stase & Romania  & Mats Wilander & Sweden  & Michael Chang & United States & Andy Roddick & United States\\ 
 \hline 
$ 4$ & Guillermo Vilas & Argentina  & Stefan Edberg & Sweden & Goran Ivani\v{s}evi\'c  & Croatia & Lleyton Hewitt & Australia\\ 
 \hline 
$ 5$ & Arthur Ashe & United States & Jimmy Connors & United States & Yevgeny Kafelnikov & Russian Federation & Nikolay Davydenko & Russian Federation\\ 
 \hline 
$ 6$ & Brian Gottfried & United States & Boris Becker & Germany & Jim Courier & United States & Ivan Ljubi\v{c}i\'c & Croatia\\ 
 \hline 
$ 7$ & Manuel Orantes & Spain & Andr\'es G\'omez & Ecuador & Richard Krajicek & Netherlands & Juan Carlos Ferrero & Spain\\ 
 \hline 
$ 8$ & Eddie Dibbs & United States & Yannick Noah & France & Thomas Muster & Austria & Novak Djokovi\'c & Serbia\\ 
 \hline 
$ 9$ & Harold Solomon & United States & Brad Gilbert & United States & Wayne Ferreira & South Africa & David Nalbandian & Argentina\\ 
 \hline 
$ 10$ & Stan Smith & United States & Tom\'a\v{s} \v{S}m\'id & Czech Republic & Thomas Enqvist & Sweden & Tommy Robredo & Spain \\ 
 \hline 
$ 11$ & Roscoe Tanner & United States & Henri Leconte & France & Boris Becker & Germany & David Ferrer & Spain\\ 
 \hline 
$ 12$ & Ra\'ul Ram\'irez & Mexico  & Tim Mayotte & United States & Stefan Edberg & Sweden & Fernando Gonz\'alez & Chile\\ 
 \hline 
$ 13$ & Tom Okker & Netherlands & Anders Jarryd & Sweden  & Sergi Bruguera & Spain & Andy Murray & Great Britain\\ 
 \hline 
$ 14$ & John Alexander & Australia & Miloslav Me\v{c}\'i\v{r} Sr. & Slovakia  & Marc Rosset & Switzerland & Carlos Moy\'a & Spain\\ 
 \hline 
$ 15$ & Vitas Gerulaitis & United States  & Kevin Curren & United States & Petr Korda & Czech Republic & Mikhail Youzhny & Russian Federation\\ 
 \hline 
$ 16$ & Ken Rosewall & Australia  & Aaron Krickstein & United States & Todd Martin & United States & James Blake & United States\\ 
 \hline 
$ 17$ & John Newcombe & Australia  & Guillermo Vilas & Argentina & C\'edric Pioline & France & Tommy Haas & United States\\ 
 \hline 
$ 18$ & Wojtek Fibak & Poland  & Joakim Nystrom & Sweden & Michael Stich & Germany & Fernando Verdasco & Spain\\ 
 \hline 
$ 19$ & Dick Stockton & United States  & Emilio S\'anchez & Spain & \`Alex Corretja & Spain & Marat Safin & Russian Federation\\ 
 \hline 
$ 20$ & John McEnroe & United States & Johan Kriek & United States & Patrick Rafter & Australia & Tom\'a\u{s} Berdych & Czech Republic\\ 
 \hline 
$ 21$ & Adriano Panatta & Italy & Martin Jaite & Argentina & Magnus Gustafsson & Sweden & Juan Ignacio Chela & Argentina\\ 
 \hline 
$ 22$ & Jan Kode\v{s} & Czech Republic  & Jakob Hlasek & Switzerland  & Andrei Medvedev & Ukraine & Radek \v{S}t\v{e}p\'anek & Czech Republic\\ 
 \hline 
$ 23$ & Jaime Fillol Sr. & Chile  & Jimmy Arias & United States & Francisco Clavet & Spain & Andre Agassi & United States\\ 
 \hline 
$ 24$ & Robert Lutz & United States & Pat Cash & Australia & Marcelo R\'ios & Chile  & Robin S\"oderling & Sweden\\ 
 \hline 
$ 25$ & Marty Riessen & United States & Ramesh Krishnan & India & Greg Rusedski & Great Britain & Rainer Sch\"uttler & Germany\\ 
 \hline 
$ 26$ & Rod Laver & Australia & Jos\'e-Luis Clerc & Argentina & Fabrice Santoro & France & Feliciano L\'opez & Spain\\ 
 \hline 
$ 27$ & Tom Gorman & United States & Eliot Teltscher & United States & Magnus Larsson & Sweden & Tim Henman & Great Britain \\ 
 \hline 
$ 28$ & Vijay Amritraj & India & Thierry Tulasne & France & Tim Henman & Great Britain & Jarkko Nieminen & Finland\\ 
 \hline 
$ 29$ & Mark Cox & Great Britain & Scott Davis & United States & Alberto Berasategui & Spain & Mardy Fish & United States\\ 
 \hline 
$ 30$ & Onny Parun & New Zealand & Vitas Gerulaitis & United States & Albert Costa & Spain & Gast\'on Gaudio & Argentina\\ 
\hline
\hline
\end{tabular}
\end{center}
\caption{Top $30$ players per decade.}
\label{table3}
\end{table*}
\end{turnpage}


\begin{turnpage}
\begin{table*}
\begin{center}
\begin{tabular}{l |l l|l l|l l|}
\cline{2-7}
& \multicolumn{2}{|c|}{{\normalsize Clay}} & \multicolumn{2}{|c|}{{\normalsize Grass}} & \multicolumn{2}{|c|}{{\normalsize Hard}}\\
Rank & Player & Country & Player & Country & Player & Country\\
\hline\hline
$ 1$ & Guillermo Vilas & Argentina  & Jimmy Connors & United States & Andre Agassi & United States\\ 
 \hline 
$ 2$ & Manuel Orantes & Spain & Boris Becker & Germany & Jimmy Connors & United States \\ 
 \hline 
$ 3$ & Thomas Muster & Austria & Roger Federer & Switzerland & Ivan Lendl & United States\\ 
 \hline 
$ 4$ & Ivan Lendl & United States & John Newcombe & Australia & Pete Sampras & United States\\ 
 \hline 
$ 5$ & Carlos Moy\'a & Spain & John McEnroe & United States & Roger Federer & Switzerland\\ 
 \hline 
$ 6$ & Eddie Dibbs & United States & Pete Sampras & United States & Stefan Edberg & Sweden\\ 
 \hline 
$ 7$ & Jos\'e Higueras & Spain & Tony Roche & Australia & Michael Chang & United States\\ 
 \hline 
$ 8$ & Bj\"orn Borg & Sweden & Stefan Edberg & Sweden & John McEnroe & United States\\ 
 \hline 
$ 9$ & Ilie N\u{a}stase & Romania & Roscoe Tanner & United States & Andy Roddick & United States \\ 
 \hline 
$ 10$ & Andr\'es G\'omez & Ecuador & Lleyton Hewitt & Australia  & Lleyton Hewitt & Australia\\ 
 \hline 
$ 11$ & \`Alex Corretja & Spain & Ken Rosewall & Australia & Brad Gilbert & United States\\ 
 \hline 
$ 12$ & Rafael Nadal & Spain & Arthur Ashe & United States & Jim Courier & United States\\ 
 \hline 
$ 13$ & Jos\'e-Luis Clerc & Argentina  & Stan Smith & United States & Brian Gottfried & United States \\ 
 \hline 
$ 14$ & Sergi Bruguera & Spain & Phil Dent & Australia & Thomas Enqvist & Sweden\\ 
 \hline 
$ 15$ & Mats Wilander & Sweden & Bj\"orn Borg & Sweden & Stan Smith & United States\\ 
 \hline 
$ 16$ & Albert Costa & Spain & Goran Ivani\v{s}evi\'c & Croatia & Boris Becker & Germany\\ 
 \hline 
$ 17$ & Gast\'on Gaudio & Argentina & Pat Cash & Australia & Wayne Ferreira & South Africa\\ 
 \hline 
$ 18$ & Juan Carlos Ferrero & Spain & Andy Roddick & United States & Ilie N\u{a}stase & Romania\\ 
 \hline 
$ 19$ & Harold Solomon & United States  & Ivan Lendl & United States & Roscoe Tanner & United States\\ 
 \hline 
$ 20$ & Emilio S\'anchez & Spain & Tim Henman & Great Britain & Tommy Haas & United States\\ 
 \hline 
$ 21$ & Adriano Panatta & Italy & Rod Laver & Australia & Rafael Nadal & Spain\\ 
 \hline 
$ 22$ & F\'elix Mantilla & Spain & Mark Edmondson & Australia & Tim Henman & Great Britain\\ 
 \hline 
$ 23$ & Francisco Clavet & Spain & John Alexander & Australia & Mats Wilander & Sweden\\ 
 \hline 
$ 24$ & Bal\'azs Tar\'oczy & Hungary & Hank Pfister & United States & Yevgeny Kafelnikov & Russian Federation\\ 
 \hline 
$ 25$ & \v{Z}eljko Franulovi\'c & Croatia & Wally Masur & Australia & Andy Murray & Great Britain\\ 
 \hline 
$ 26$ & Tom\'a\v{s} \v{S}m\'id & Czech Republic & Tim Mayotte & United States & Fabrice Santoro & France\\ 
 \hline 
$ 27$ & Jimmy Connors & United States & Vijay Amritraj & India & Harold Solomon & United States\\ 
 \hline 
$ 28$ & Ra\'ul Ramirez & Mexico & Kevin Curren & United States & Ivan Ljubi\v{c}i\'c & Croatia\\ 
 \hline 
$ 29$ & Alberto Berasategui & Spain & Tom Okker & Netherlands & Marat Safin & Russian Federation\\ 
 \hline 
$ 30$ & Victor Pecci Sr. & Paraguay & Greg Rusedski & Great Britain & Aaron Krickstein & United States\\ 
\hline
\hline
\end{tabular}
\end{center}
\caption{Top $30$ players of the history
of tennis in tournaments played on clay, grass and hard surfaces.}
\label{table4}
\end{table*}
\end{turnpage}



\end{document}